\documentclass[aps,preprint,floats,prb]{revtex4}
\usepackage{graphics,dcolumn}

\begin{document}

\title{First-principles molecular dynamics simulations at solid-liquid 
interfaces with a continuum solvent}
\author{Ver\'{o}nica M. S\'{a}nchez$^\dag$}
\author{Mariela Sued$^\ddag$}
\author{Dami\'{a}n A. Scherlis$^\dag$}
\affiliation{$^\dag$Departamento de Qu\'{i}mica Inorg\'{a}nica, Anal\'{i}tica
y Qu\'{i}mica F\'{i}sica/INQUIMAE, Facultad de Ciencias Exactas
y Naturales, Universidad de Buenos Aires, Ciudad Universitaria,
Pab. II, Buenos Aires (C1428EHA) Argentina}
\affiliation{$^\ddag$Instituto de C\'{a}lculo,
Facultad de Ciencias Exactas
y Naturales, Universidad de Buenos Aires, Ciudad Universitaria,
Pab. II, Buenos Aires (C1428EHA) Argentina}

\begin{abstract}

Continuum solvent models have become a standard technique
in the context of electronic structure calculations, yet, no implementations
have been reported capable to perform molecular dynamics at solid-liquid interfaces.
We propose here such a continuum approach in a DFT framework, using plane-waves basis sets
and periodic boundary conditions. Our work stems from a recent model 
designed for Car-Parrinello simulations of quantum solutes in a dielectric medium
[J. Chem. Phys. {\bf 124}, 74103 (2006)], for
which the permittivity of the solvent is defined as a function of the electronic 
density of the solute. This strategy turns out to be inadequate for systems extended in two dimensions:
the dependence of the dielectric function 
on the electronic density introduces a new term in the Kohn-Sham potential
which becomes unphysically large at the interfacial region,
seriously affecting the convergence of the self-consistent calculations.
If the dielectric medium is properly redefined as a function of
the atomic coordinates, a good convergence is obtained and the constant of motion
is conserved during the molecular dynamics simulations.
Moreover, a significant gain in efficiency can be achieved if the simulation box is partitioned
in two, solving the Poisson problem separately for the ``dry'' region using fast Fourier transforms,
and for the solvated or ``wet'' region using a multigrid method. Eventually
both solutions are combined in a self-consistent procedure, and in this way
Car-Parrinello molecular dynamics simulations
of solid-liquid interfaces can be performed at a very moderate computational cost.
This scheme is employed to investigate the acid-base equilibrium at the TiO$_2$-water interface.
The aqueous behavior of titania surfaces has stimulated a large amount of experimental research,
but many open questions remain concerning the molecular mechanisms determining the
chemistry of the interface. Here we make an attempt to answer some of them,
putting to the test our continuum model.

\end{abstract}

\date{\today}
\maketitle

\section{Introduction}

The structure and the reactivity of solid surfaces have become major subjects
of study in chemistry and condensed matter physics, and are at the core of
much of the research conducted in materials science. In this context, electronic structure
methods, in particular density functional theory (DFT) calculations, have played
a fundamental role in the interpretation and the validation of the data
coming from the different surface spectroscopies and microscopies, often providing
new insights on top of those detaching from the available experimental techniques.\cite{science}
Throughout the past two decades a large number of DFT simulations has been reported 
on metallic and semiconducting surfaces, contributing precious information concerning
atomic and electronic structure, thermodynamics, and reactivity.\cite{revsurf}$^-$\cite{diebold}
With very few exceptions, these simulations
considered a slab in the gas phase. For a broad range of applications, however, the relevant
phenomena occur in the presence of a liquid phase, as is often the case in processes related to
electrochemistry and catalysis. The realization of liquid phase DFT simulations is therefore 
a much pursued objective---especially when many of the standard
X-ray techniques like XPS or SAXS are unsuited to provide atomic scale information in solution---,
but the inclusion of the
solvent considerably increases the cost of first-principles calculations of periodic 
surfaces, and is therefore a rather uncommon practice.\cite{examples}

In an explicit solvation approach, the vacuum space in the simulation box is filled with solvent molecules.
The subsequent growth in the size of the system is in part responsible for the increased computational expense,
but, more importantly, there is the fact that a static picture is a very poor representation
of the liquid state. Any solute admits a huge number of possible configurations for
the solvent molecules around it, associated with multiple local minima, the net solvation effect arising
from a weighted average of all these.\cite{crtruhlar}
A geometry optimization would lead to a solid or glassy phase
corresponding to one of these minima in the multidimensional potential energy
surface, resulting in a dielectric screening typically different from the static limit
observed in the liquid state. Hence, to capture the solvation effect, it is necessary
to perform extensive statistical sampling involving
either lengthy molecular dynamics or Monte-Carlo simulations.
Alternatively, it is possible to resort to the so called continuum (or implicit solvent) models,
in which the solvent molecules are replaced by a dielectric medium surrounding the solute 
and exhibiting the static screening of the solution.\cite{levine,tamar}
In this way, the polarization induced by the solvent is introduced in an averaged fashion,
and the cost of the computation gets closer to the corresponding cost in vacuum by drastically reducing the
number of degrees of freedom. On the other hand, the representation of the solvent structure is omitted, 
disregarding any possible solute-solvent specific interactions. 
Still, to overcome this problem, all or part of the first solvation shells can
be included explicitly, with the dielectric medium extending beyond the limits of this
cluster comprising the solute plus a few solvent molecules. 
In any case, the continuum model has a long tradition in quantum chemistry and
has proved reliable and efficient to extract properties in solution of a large variety of
molecular systems.\cite{crtruhlar}$^-$\cite{crtomasi1}

In recent years a small number of implementations of the continuum model has been
proposed in the context of density functional theory, plane-waves basis sets, and the Car-Parrinello
method.\cite{cpldangelis}$^-$\cite{jcp0} 
To the best of our knowledge, none of them have been employed in molecular
dynamics simulations of periodic surfaces.\cite{arias}
Here we revise the electrostatic continuum model of Fattebert and Gygi,\cite{jluc2,jcp0,jluc1} 
proposing a new definition for the dielectric function to make the model suitable for the treatment 
of periodic slabs.
The original formulation, in which the permittivity of the solvent is determined 
by the charge density of the solute, has proved successful in the simulation
of molecular and ionic solutes,\cite{jluc2,jcp0} and of extended systems like polymers, with periodicity
in one dimension.\cite{jcp1} 
In the case of solid surfaces, however, the convergence to the electronic ground state turns out
to be impaired. In the sections that follow, we attribute this failure to a term
in the Kohn-Sham potential originating in the dependence of the dielectric constant on the
electronic density, and, to circumvent this issue, we redefine the permittivity
as a function of the atomic positions. In this way the smooth electronic convergence is restored,
along with a good conservation of the total energy in the molecular dynamics runs.
The electrostatic problem in the presence of the dielectric medium is efficiently addressed with a multigrid
method,\cite{nrecipes}$^-$\cite{mgmethods} which solves the Poisson equation in real space.
Moreover, a significant increase in efficiency can be obtained if the simulation box is partitioned
in two, solving the Poisson problem separately for the ``dry'' region using fast Fourier transforms,
while restricting the multigrid treatment to the solvated or ``wet'' part of the supercell (and
then combining both solutions in a self-consistent fashion).
Using this scheme it is possible to carry out Car-Parrinello molecular dynamics simulations 
of solid-liquid interfaces at a computational cost exceeding by just
a small factor the one corresponding to vacuum.
We employ this approach to investigate the proton exchange at the
anatase-water interface, which is a key process in the acid-base equilibrium of TiO$_2$ surfaces.
The ubiquity of titania in solid-liquid applications has incited the
emergence of empirical models to assess the protonation and the dissociation of terminal
groups on the TiO$_2$ surface in aqueous environments.\cite{music,jolivet} 
These simple models have been broadly used among experimentalists for the
interpretation of their data, but a molecular level description that accounts for the effect
of structure has not been established.
In this context, the present simulations intend to provide a first-principles 
glance of the microscopic mechanisms governing the chemistry of this interface.

\section{Model and methodological discussion}

\subsection{About the present implementation and the computational parameters}

This model has been implemented in the
public domain Car-Parrinello parallel code included in the Quantum-ESPRESSO
package,\cite{espresso} based on density-functional theory (DFT),
periodic-boundary conditions, plane-waves basis sets, and
pseudopotentials to represent the ion-electron interactions.
All calculations reported in this work, unless otherwise noted,
have been performed using the PW91 exchange-correlation functional\cite{PW91}
in combination with Vanderbilt ultrasoft pseudopotentials.\cite{usp} 
The Kohn-Sham orbitals and charge density were expanded in plane waves 
up to a kinetic energy cutoff of 25 and 200 Ry respectively.
Periodic slabs of four layers width representing the (101) surface of the anatase structure
were computed using gamma-point sampling in supercells of
size 10.24 $\AA$ x 7.56 $\AA$ x 17.82 $\AA$.
For finite temperature (not damped) Car-Parrinello molecular dynamics simulations, an
electronic mass of 400 a.u. and a time step of 0.17 fs were adopted.
During geometry relaxations and dynamics, all atoms were allowed to move, with the
exception of those belonging to the two inner layers, which were fixed in their bulk positions.

\subsection{The continuum solvation model}

Within the continuum approach, the solvent is represented as
a dielectric medium surrounding a quantum-mechanical
solute confined in a cavity. In particular, we consider the polarizable continuum model,
in which the dielectric medium and the electronic density
respond to the field of each other in a self-consistent
fashion. This interaction provides the electrostatic part of the solvation
free energy, $\Delta G_{el}$, which is the dominant contribution for
polar and charged solutes. The cavitation energy $\Delta G_{cav}$ is defined as the work involved
in creating the appropriate cavity inside the solution in the
absence of solute-solvent interactions.\cite{levine}
Electrostatic, cavitation, and
dispersion-repulsion effects are modeled as separate contributions,\cite{levine} and the 
free energy of solvation is regarded as the sum of these three terms
($\Delta G_{sol} = \Delta G_{el}+\Delta G_{cav}+\Delta G_{dis-rep}$). Thermal
and pressure dependent terms can also be included but are usually negligible.
We note that this decomposition is inherent to the model, being $\Delta G_{sol}$
the only measurable quantity.

\subsection{Previous implementation}

The starting point for this work is the implementation reported in reference~~\cite{jcp0}:
in the following paragraphs we revisit those aspects of the preceding version
that are essential to the present development. First, we note
$\Delta G_{el}$ and $\Delta G_{cav}$ are considered explicitly,
while $\Delta G_{dis-rep}$, less relevant for solutes of moderate size,
is largely seized as part of the electrostatic term by virtue of the parametrization.
The electrostatic interaction between the dielectric and the solute
is calculated as proposed by Fattebert and Gygi,\cite{jluc2,jluc1} who define
the permittivity $\epsilon$ of the solvent as a function of the electronic density $\rho$.
Within a pseudopotential framework and in periodic boundary conditions,
the Kohn-Sham energy functional\cite{marx-hutter} can be written as
\[
E[\rho] = T[\rho]+ E_{xc}[\rho] + \frac{1}{2} \int \phi({\bf r}) \rho_{tot}({\bf r})d{\bf r} \]
\begin{equation}
+\sum_{I<J} \frac{Z_I Z_J}{R_{IJ}} erfc \biggl(
\frac{R_{IJ}}{\sqrt{(R_I^c)^2+(R_J^c)^2}} \biggr)  
-\frac{1}{\sqrt{2\pi}} \sum_I \frac{Z_I^2}{R_{I}^c} + E_{ps}[\rho]
\end{equation}
where $T[\rho]$ corresponds to
the kinetic energy of the electrons and $E_{xc}[\rho]$ to the exchange-correlation energy.
The last four terms on the right hand side account for the total electrostatic energy in a periodic
crystal, gathering all Coulombic interactions involving electrons and nuclei, only
omitting for simplicity the non-local part of the pseudopotential
(for a detailed derivation see reference~\cite{galli} or the Appendix of reference~\cite{jcp0}).
The electrostatic formulation in Eq. (1) arises from the Ewald sum of point charges, which 
requires to introduce the ionic densities $\rho_I({\bf r-R_I})$ consisting of
Gaussian distributions of negative
sign that integrate to $Z_I$, the total charge of the pseudo-ion:
$\rho_I({\bf r-R_I})=-\frac{Z_I}{(R_{I}^c)^3} \pi^{-\frac{3}{2}}
\exp \biggl(-\frac{|r-R_I|^2}{(R_{I}^c)^2}\biggr),$
with $R_{I}^c$ the width of the Gaussian associated with
the site $I$. 

The third term on the right is conventionally called the Hartree energy:
\[ E_H = \frac{1}{2} \int \phi({\bf r}) \rho_{tot}({\bf r})d{\bf r} \]
where $\rho_{tot}$ is just the sum of the electronic plus
the ionic densities, $\rho_{tot}({\bf r})=\rho({\bf r})+\sum_I \rho_I({\bf r-R_I})$,
while the electrostatic potential $\phi[\rho]$ is the
solution to the Poisson equation in vacuum,
\begin{equation}
\nabla ^2 \phi({\bf r}) = -4\pi \rho_{tot}({\bf r})~.
\end{equation}
In the presence of a dielectric continuum with a permittivity
$\epsilon[\rho]$, the Poisson equation becomes
\begin{equation}
\nabla \cdot (\epsilon[\rho] \nabla \phi({\bf r})) = -4\pi \rho_{tot}({\bf r})~.
\end{equation}
Using Eq.~(3), the formula for the Hartree energy $E_H$, can be integrated by parts to yield:
\begin{equation}
E_H = \frac{1}{8\pi} \int \epsilon[\rho](\nabla \phi({\bf r}))^2d{\bf r}.
\end{equation}
The functional derivative of $E_H$ with respect to the charge density is added to the other
contributions of the energy (the exchange-correlation,
the local and non-local parts of the pseudopotentials) to set up the Kohn-Sham potential $V^{KS}[\rho]$.
\begin{equation}
\frac{\delta E_H}{\delta \rho}({\bf r}) = 
\phi({\bf r}) + V_{\epsilon}({\bf r}),
\end{equation}
\begin{equation}
V_{\epsilon}({\bf r})=-\frac{1}{8\pi} (\nabla \phi({\bf r}))^2 
\frac{\delta \epsilon}{\delta \rho}(\bf r).
\end{equation}
We provide the derivation of Eq.~(5) and (6) in the Appendix A, since it is not given in any of the
previous references. The dielectric medium and the
electronic density respond self-consistently to each other
through the dependence
of $\epsilon$ on $\rho$ and vice versa.

In quantum chemistry continuum models as PCM,\cite{crtomasi1,cpl-cossi}
the dielectric constant $\epsilon$ is
taken to be 1 inside the cavity, and a fixed value outside.
For molecular dynamics applications, such a discontinuity needs to be removed to calculate accurately
the analytic derivatives of the potential with respect to the
ionic positions. Besides, in the particular case of plane-waves implementations based on real space
grids, a smoothly varying dielectric function is more appropriate for numerical reasons.
Fattebert and Gygi proposed the following smoothed step function for the dielectric:
\begin{equation}
\epsilon(\rho({\bf r})) = 1+ \frac{\epsilon_\infty -1}{2}\left (
1+\frac{1-(\rho({\bf r})/\rho_0)^{2\beta}}{1+(\rho({\bf r})/\rho_0)^{2\beta}}
\right ).
\end{equation}
This function asymptotically approaches $\epsilon_\infty$
(the permittivity of the bulk solvent) in regions of space where the
electron density is low, and 1 in those regions where it is high (outside the solvation cavity).
The parameter $\rho_0$ is the density threshold determining the cavity
size, whereas $\beta$ modulates the smoothness of the transition from
$\epsilon_\infty$ to 1.

On the other hand, the cavitation energy is computed separately as the product between the surface
tension of the solvent and the area of the cavity. In this implementation this term
remains unchanged; details about the calculation of $\Delta G_{cav}$ can be 
found in reference~~\cite{jcp0}.

\subsection{Calculation of periodic slabs: the problem in the convergence}

\begin{figure} \centerline{
\rotatebox{-90}{\resizebox{3.5in}{!}{\includegraphics{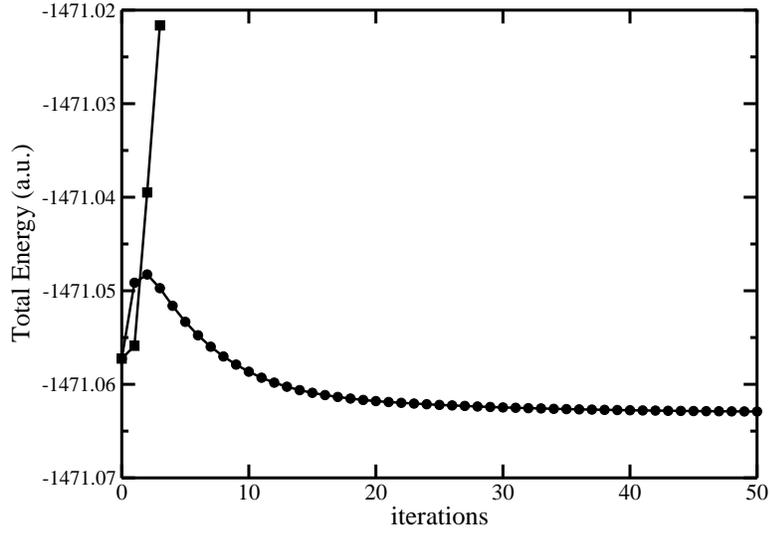}}}
}
\caption{
Total energy during two different Car-Parrinello electronic minimizations in a
continuum solvent (the structure is the TiO$_2$ slab described in section III).
The squares indicate the results for
a dielectric function depending on the charge density, whereas the circles correspond to the same
computational experiment, but removing $V_\epsilon$ from the total potential
(see text). Both runs used the same calculation parameters and were restarted from
the wavefunction converged in vacuum.}
\label{KineticE}
\end{figure}

When the scheme described above is applied on structures extended in two dimensions, it typically
fails to achieve self-consistency. We found the electrons heat up during the
Car-Parrinello relaxation of the wavefunction, causing the total energy to diverge,
as shown in Fig.~\ref{KineticE}.
In other cases the kinetic energy of the electrons is observed to decrease at the beginning,
but thereafter experiences irregular oscillations without ever reaching zero.
Even if convergence is enforced through the use
of a stringent algorithmic strategy, the solution obtained is not reliable, e.g.,
it depends on the starting conditions or the damping parameters.

\begin{figure} \centerline{
\rotatebox{-90}{\resizebox{3.5in}{!}{\includegraphics{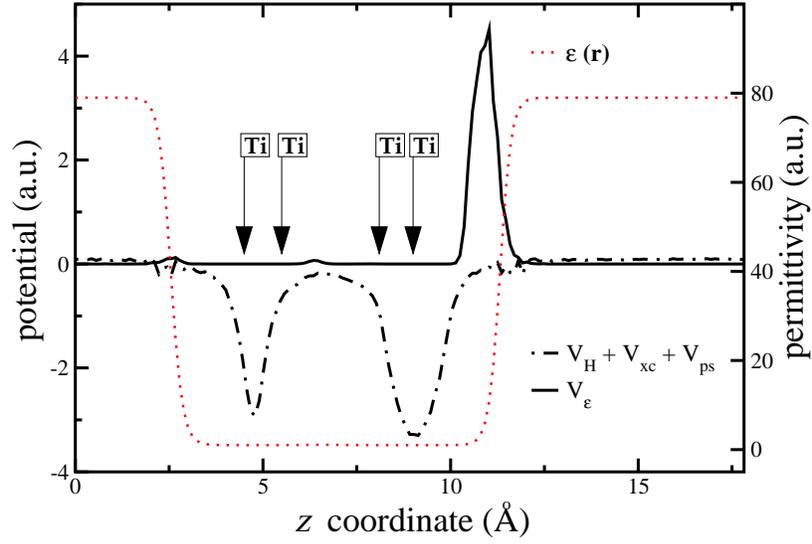}}}
}
\caption{
Variation of $V_\epsilon$ and of the effective potential along the $z$-direction (perpendicular to the
slab) at a selected point on the $x$-$y$ plane, corresponding to the position of one of the exposed Ti
atoms of the surface. The dotted line represents the permittivity,
referenced on the right axis. The arrows point to the approximate location of each of the titanium layers.}
\label{vepsilon2D}
\end{figure}

The source of this erratic behavior can be tracked to the inclusion of the term $V_\epsilon$
in the Kohn-Sham potential. This term, defined in Eq.~(6), arises from the dependence
of the permittivity on the electron density. Fig.~\ref{KineticE} illustrates the impact
of this contribution on the convergence of a Car-Parrinello electronic relaxation, by
showing that an acceptable dynamics is recovered if $V_\epsilon$
is neglected.\cite{nota1} In particular, these results correspond to the anatase-water structure
described in section III, but the convergence of other systems is degraded in a similar way.
The reason for the instability associated with $V_\epsilon$ can be somehow
appreciated in the two dimensional plot of Fig.~\ref{vepsilon2D}, which displays the variation of
$V_\epsilon$ in the $z$-direction (the one perpendicular to the surface) at fixed $x$ and $y$.
These coordinates have been chosen in coincide with the position of a Ti atom of the surface.
The same graph shows the total Kohn-Sham potential, excluding $V_\epsilon$, 
so that the relative contribution of this term
can be clearly seen. The values are exceptionally large at the solid-liquid boundary, even
larger than any of the other contributions to the effective potential. Referenced on the right
axis, the dielectric function $\epsilon[\rho]$ is also shown. Since $V_\epsilon$ depends on
$\partial \epsilon / \partial \rho$ (Eq.~6), the peaks occur at the
region where the charge density decays abruptly, producing a rapid variation in $\epsilon[\rho]$ 
from the value in the solid to the value in the
solvent. Adopting the model function of Eq.~(7) we have
\begin{equation}
\frac{\partial \epsilon}{\partial \rho}({\bf r}) = \frac{1-\epsilon_\infty}{\rho_0}
\frac{2\beta(\rho({\bf r})/\rho_0)^{2\beta-1}}{(1+(\rho({\bf r})/\rho_0)^{2\beta})^2}.
\end{equation}
This expression goes to zero when $\rho({\bf r})\gg\rho_0$ or
$\rho({\bf r})\ll\rho_0$, and is dominated by $1/\rho_0$ otherwise.\cite{nota2} Then, the extreme values
of $V_\epsilon$ at the solid-liquid interface ultimately originate in the magnitude of
the density threshold $\rho_0$. Unfortunately, this parameter may not be freely tuned, but
is set in combination with $\beta$ to fit the experimental data, and no reasonable 
choice of ($\rho_0, \beta$) can be made as to prevent the blowup of $V_\epsilon$.
Alternative model functions to represent the dielectric were considered, e.g., with $\epsilon[\rho]$
exhibiting a Gaussian or trigonometric decay with $\rho$. None of these functions entailed any
significant improvement, as far as all of them have in common a sudden change
associated with the transition from 1 to $\epsilon_\infty$, which redounds
in large values of $\partial \epsilon / \partial \rho$ at the interface.
The transition can be smoothed enough as to avoid the sharp 
peaks in $V_\epsilon$, but in doing so the solvation effect is ruined.
Clearly, the discussed behavior is a consequence of the dependence of $\epsilon$ on $\rho$,
irrespective of the kind of function chosen to model the dielectric.
It should be noticed that the inclusion of $V_\epsilon$ does not appear to 
disturb the convergence or the energy conservation in the case of finite systems, neither
in the case of polymers (extended in one dimension).\cite{jcp1} It is seemingly because of the
bi-dimensional symmetry of periodic surfaces that $V_\epsilon$ turns out to spoil
the Car-Parrinello dynamics, by propagating the observed perturbation of the Kohn-Sham
potential throughout a full, extended plane.

\subsection{A position dependent formulation}

It is fair to wonder about the physical meaning of this strong perturbation of the potential
contained in $V_\epsilon$. In the absence of a dielectric, or for a dielectric defined independently
of the charge density, the functional derivative of $E_{H}$ with respect to $\rho$ turns out to be
equal to the electrostatic potential $\phi[\rho]$ (see reference \cite{parryang}). 
The additional term $V_\epsilon$ emerges from the dependence of $E_{H}$ on $\epsilon$, whenever
$\epsilon$ is modeled as a function of the charge density. Since $\partial \epsilon / \partial \rho < 0$
(in Eq. (8), $\epsilon_\infty>1$), $V_\epsilon$ is of a repulsive character.
The choice of $\rho$ to delimitate the region filled by the continuum induces a response of the potential 
to counteract the strong discontinuity imposed by the same $\rho$.
$V_\epsilon$ is such a response, which, throughout the self-consistent procedure, opposes
the perturbative drift enforced by the model.

Thus, the instability of the model emanates from the use of the self-consistent charge density to
define the dielectric. A possible alternative would be to use instead a non-self
consistent, or ``fake'' density, since the role of the charge density in this context is simply to shape
the dielectric medium. As will be discussed below, such an option can be equivalent to define
$\epsilon$ as a function of the atomic coordinates, which is the usual strategy
in quantum chemistry methods. The idea of a dielectric determined by the atomic positions
may be less attractive from a physical viewpoint: the size and shape of the cavity depend on the identity
of the atoms only, and is not modulated by the electronic structure or the environment; the polarization of
the solvent is lost, and, on top of these, many more
parameters are needed---at least one per atom. In practice, however, it is possible
to choose a dielectric function based on the molecular coordinates that closely
reproduces the $\rho$-dependent solvation, because the effect of the self-consistency and
the polarization of the solvent on $\Delta G_{sol}$ is quite minor.

In this way, we keep the expression for $\epsilon$ given in Eq. (7), but feed it with a dummy
density $\gamma({\bf r})$ determined by the ionic positions ${\bf R_I}$,
\begin{equation}
\gamma({\bf r})=\sum_I e^{-\left(|{\bf r - R_I}|-R^I_{vdw}\right)},
\end{equation}
where $R^I_{vdw}$ is the van der Waals radius for the corresponding species. 
Hence the dielectric function takes the following form:
\begin{equation}
\epsilon(\gamma({\bf r}))=1+ \frac{\epsilon_\infty -1}{2}\left (
1+\frac{1-\gamma({\bf r})^{2\beta}}{1+\gamma({\bf r})^{2\beta}}
\right ).
\end{equation}

\begin{figure} \centerline{
\rotatebox{-90}{\resizebox{3.5in}{!}{\includegraphics{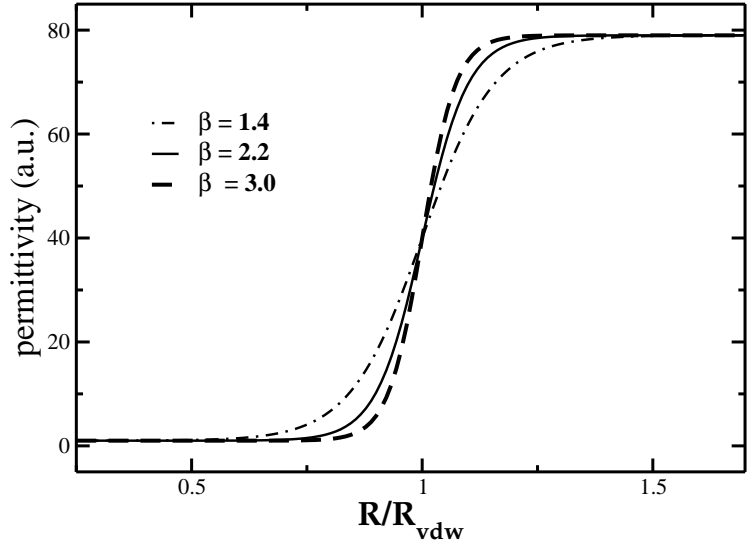}}}
}
\caption{The permittivity around an oxygen atom as a function of the distance, for different values of
the parameter $\beta$, according to the position-dependent dielectric defined
in Eqs. (9) and (10). The transition, centered around the van der Waals radius, becomes sharper
as $\beta$ is increased.}
\label{gamma}
\end{figure}

Using this definition, the transition of $\epsilon(\gamma({\bf r}))$ between 1 and $\epsilon_\infty$
is centered around the van der Waals radius. Aside from $R^I_{vdw}$, which values
are tabulated, $\beta$ is left as the only
adjustable parameter to fit the experimental
solvation energies. Fig.~\ref{gamma} shows the aspect of the dielectric function around
an oxygen atom for different $\beta$. This parameter must be large enough to ensure most of the
transition occurs within a small length window, but at the same time the
numerical accuracy needs to be preserved, so there is an upper bound for $\beta$
which depends on the given grid size.

Within this framework, the electrostatic contribution to $V^{KS}[\rho]$ is simply the electrostatic
potential $\phi[\rho]$, and the stability of Car-Parrinello dynamics in periodic slabs
is recovered.
Table I presents, for several neutral and charged solutes, a comparison between the values of
$\Delta G_{sol}$ obtained with the dielectric functions of Eq.~(7) and Eq.~(10), respectively.
The small differences proceed exclusively from $\Delta G_{el}$, since
$\Delta G_{cav}$ is the same in both cases ($\Delta G_{sol} = \Delta G_{el} + \Delta G_{cav}$).
As mentioned above, with a proper choice of $\beta$ the position dependent dielectric is able
to provide solvation energies in close agreement with the previous model and with experiments.
The results shown correspond to $\beta = 2.4$. 

The explicit dependence of $\epsilon$ on the ionic positions involves a new contribution to
the forces arising from the derivative of $E_{H}$ with respect to ${\bf R_I}$, which must
be taken into account to perform conservative molecular dynamics simulations. After some
manipulation involving the use of Eqs. (3) and (4),
this derivative can be expressed as follows (the full derivation is given in Appendix B): 
\begin{equation}
\frac{\partial E_H}{\partial {\bf R_I}}
= -\frac{1}{8\pi} \int \frac{\partial \epsilon({\bf R_I})}
{\partial {\bf R_I}}(\nabla \phi({\bf r}))^2d{\bf r}
+ \int \phi({\bf r}) \frac{\partial \rho_{tot}({\bf r})}{\partial {\bf R_I}}d{\bf r}
\end{equation}
The computation of the first term on the right is straightforward since we know, from Eqs. (9) and (10),
the explicit dependence of $\epsilon$ on ${\bf R_I}$:
\begin{equation}
\frac{\partial \epsilon({\bf R_I})}{\partial \tau_I}({\bf r}) = 
2\beta \left(\epsilon_\infty -1 \right) \left(\frac{\tau-\tau_0}{R}\right) 
\frac{e^{-(R-R_{vdw}^I)} \left(\sum_I e^{-(R-R_{vdw}^I)} \right)^{2\beta-1}}
{\left[1+\left(\sum_I e^{-(R-R_{vdw}^I)} \right)^{2\beta} \right]^2}
\end{equation}
with $\tau$ a generic coordinate $x, y,$ or $z$, ${\bf R_I}=(x_0, y_0, z_0)$ and $R=|{\bf r-R_I}|$.

On the other hand, if $\phi({\bf r})$ is transformed
to Fourier space so that $\phi({\bf r})=\sum_{\bf G} \tilde \phi({\bf G}) e^{i{\bf G r}}$
(see next section), the second term in Eq. (11) can be evaluated as
\begin{equation}
\int \phi({\bf r}) \frac{\partial \rho_{tot}({\bf r})}{\partial {\bf R_I}}d{\bf r}=
-\Omega \sum_{\bf G} i{\bf G} \tilde \phi^*({\bf G}) \tilde \rho_I({\bf G}) e^{-i{\bf G R_I}}
\end{equation}
with $\tilde \rho_I({\bf G})$ the form factor of the ionic densities, 
$\rho_I({\bf r-R_I})=\sum_G \tilde \rho_I({\bf G}) e^{-i{\bf G r}} e^{-i{\bf G R_I}}$. 

To ensure the conservation of the total energy during the molecular dynamics simulations,
the contributions given in Eqs. (12) and (13) must take the place of the derivative of
the Hartree energy in the absence of the dielectric:
\begin{equation}
\frac{\partial E_H}{\partial {\bf R_I}}=
-4\pi\Omega \sum_{\bf G} i{\bf G} \left(\frac{\tilde \rho^*({\bf G})}{G^2}\right)
\tilde \rho_I({\bf G}) e^{-i{\bf G R_I}}.
\end{equation}

\subsection{The multigrid scheme and a mixed strategy to solve the Poisson problem}

In standard plane waves codes based on real space grids, the electrostatic potential $\phi(\bf r)$
is obtained from
the Poisson equation (2), which can be efficiently inverted with the use of fast Fourier transforms (FFT).
Both the total charge density $\rho_{tot}(\bf r)$ and $\phi(\bf r)$ can be expanded in plane waves,
\[ \rho_{tot}({\bf r})= \sum_{\bf G} \tilde \rho({\bf G}) e^{i{\bf G r}},~~~~~
\phi({\bf r})=\sum_{\bf G} \tilde \phi({\bf G}) e^{i{\bf G r}}. \]
Replacing into the Poisson equation $\nabla^2 \phi=-4\pi \rho_{tot}$, and equating coefficients: 
\begin{equation}
{\bf G}^2 \tilde \phi({\bf G}) = 4\pi \tilde \rho({\bf G})~~,~~~~
\phi({\bf r})= \sum_{\bf G} \frac{4\pi}{{\bf G}^2} \tilde \rho({\bf G}) e^{i{\bf G r}}
\end{equation}
Unfortunately, the Poisson equation in the presence of an arbitrary dielectric, Eq. (3), can
not be addressed in the same way, and an alternative numerical scheme is required.
To that end, we have implemented from scratch a sixth-order multigrid code,\cite{nota3}
which solves in real space the Poisson equation with non-constant coefficients and periodic 
boundary conditions. Eq. (3) can be rewritten as:
\begin{equation}
\frac{\partial \epsilon}{\partial x} \frac{\partial \phi}{\partial x} +
\frac{\partial \epsilon}{\partial y} \frac{\partial \phi}{\partial y} +
\frac{\partial \epsilon}{\partial z} \frac{\partial \phi}{\partial z} +
\epsilon \left(\frac{\partial^2 \epsilon}{\partial x^2}+
\frac{\partial^2 \epsilon}{\partial y^2}+\frac{\partial^2 \epsilon}{\partial z^2}
\right)  = -4\pi \rho.
\end{equation}
This equation is developed in finite differences, expanding the derivatives of $\phi$ and
$\epsilon$ to sixth-order according to the following relations for the gradient and
the Laplacian:
\begin{equation}
\frac{\partial f({\bf r})}{\partial \tau} = \frac{1}{h}\sum_{n=-3}^{3} \alpha_n~u_{i+n} + O(h^7)
\end{equation}
\begin{equation}
\frac{\partial^2 f({\bf r})}{\partial \tau^2}=\frac{1}{h^2}\sum_{n=-3}^{3} \beta_n~u_{i+n} + O(h^7)
\end{equation}
where $\tau$ is a generic coordinate $x, y,$ or $z$; $h$ is the grid spacing in the direction $\tau$; and
$u$ is the discretization of a continuous function $f({\bf r})$, representing $\phi({\bf r})$,
$\rho({\bf r})$, or $\epsilon({\bf r})$.
$u_i$ refers to $u$ evaluated at a mesh point associated with ${\bf r}$, while $u_{i+n}$
corresponds to a neighboring point $n$ positions to the right in the direction
$\tau$ (if $n<0, u_{i+n}$ is located to the left
of $u_i$). The coefficients $\alpha_n$ and $\beta_n$ are given by:
\[\alpha_0=0,~\alpha_1=\frac{3}{4},~ \alpha_2=-\frac{3}{20},~ \alpha_3=\frac{1}{60},~\alpha_{-n}=-\alpha_n\]
\[\beta_0=-\frac{49}{18},~\beta_1=\frac{27}{18},~ \beta_2=-\frac{27}{180},
~ \beta_3=\frac{2}{180},~\beta_{-n}=\beta_n .\]
In the case of $\epsilon({\bf r})=1$ for all ${\bf r}$, this method provides a solution 
for $\phi({\bf r})$ which is indistinguishable from the one obtained with FFT.
It also demonstrated an excellent performance when tried on functions with
non-constant coefficients and oscillation frequencies comparable to those of interest.
For example, for $\phi=e^{-ar}$ and $\epsilon=e^{-br}$ ($0.5 < a,b < 2.0$) the relative error 
in $\phi({\bf r})$ was less than 10$^{-4}$ in a mesh of 80$\times$80$\times$80 points. 

\begin{figure} \centerline{
\resizebox{5in}{!}{\includegraphics{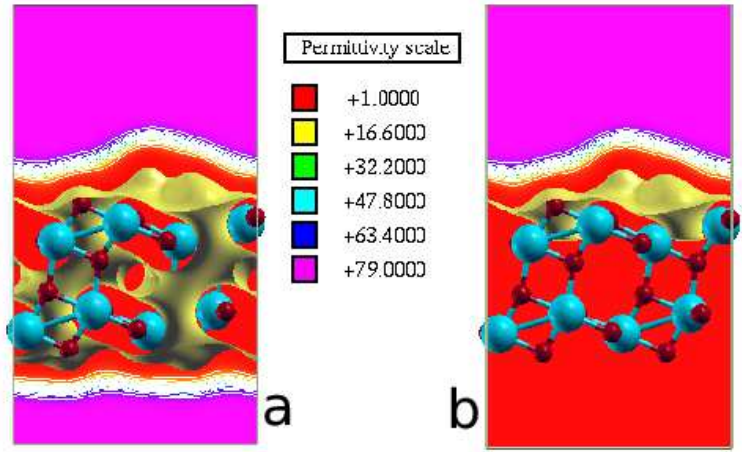}}
}
\caption{Contour plot of the dielectric function $\epsilon({\bf r})$ in a supercell containing 
a four layers slab representing the anatase (101) face of TiO$_2$.
An isosurface corresponding to $\epsilon({\bf r})$=1.4 is displayed in yellow.
In (a) the solvent percolates through the surface, whereas in (b) it is excluded from the slab by
virtue of the artifact given in relation (19).
}
\label{3Ddiel}
\end{figure}

At the initial steps of a molecular dynamics simulation, the convergence of the potential 
may require 15-30 multigrid cycles. Given the self-consistent nature of the procedure, however,
after a few time steps the number of cycles necessary to reach convergence is
typically decreased to less than five. Even so, the multigrid algorithm is still significantly more
expensive than FFTs. Propitiously, multigrid methods can be adapted to any kind of boundary
conditions, and this feature can be exploited to reduce the size of the mesh involved.
To implement this idea, a region in the supercell---preferably the slab---must remain inaccessible
to the solvent, so that $\epsilon({\bf r})=1$ within it.
In practice, the dielectric function of Eqs. (7) or (10) is not diffuse enough as to
encompass all the volume of the slab---if it were, the solvation effect would
fade at the molecular boundaries---, so the solvent occupies the interstitial space left by the atomic
structure (Fig.~\ref{3Ddiel}a). This is inconvenient not only because it
increases the numerical complexity of the problem, but also because it is not physical, i.e., the solvent
does not penetrate the atomic structure of the surface.
A simple device to exclude the solvent from the solid interspaces is to modify
$\gamma({\bf r})$ in the following way:
\begin{equation}
\gamma({\bf r}) =\left\{ \begin{array}{ll}
\sum_I e^{-\left(|{\bf r - R_I}|-R^I_{vdw}\right)} & z_I > z_{lim}\\
\sum_I e^{-\frac{z-z_I}{|z-z_I|}\left(|{\bf r - R_I}|-R^I_{vdw}\right)} & z_I \leq z_{lim}
\end{array} \right.
\end{equation}
where $z_I$ is the $z$-component of ${\bf R_I}$ (we recall $z$ is the coordinate
perpendicular to the surface; it is zero at the bottom of the unit cell and maximum at the
top). The factor $\frac{z-z_I}{|z-z_I|}$ produces a rapid increase in
$\gamma({\bf r})$ underneath the atom $I$, which saturates the value of $\epsilon({\bf r})$ for
all atoms $I$ located below $z_{lim}$. Thus, only the upper face of the slab is in contact with the
solution, the dielectric function becoming equal to 1 throughout the lower section of the supercell.
Fig.~\ref{3Ddiel}b depicts $\epsilon({\bf r})$ when $z_{lim}$ is chosen equal to the $z$ 
coordinate of an ion belonging to the second layer.

Under these conditions, it is possible to solve the Poisson problem in two steps: first,
the potential $\phi({\bf r})$ is computed for $\epsilon=1$ in the full domain $\Omega$ using FFTs;
secondly, $\phi({\bf r})$ is recalculated in the presence of the dielectric with the multigrid
method but in a smaller domain $\Omega_1$, imposing the boundary conditions proceeding from the
FFT solution. The sub-domain $\Omega_1$ must be chosen in such a way that it overlaps
with a sub-space where $\epsilon({\bf r})=1$, so that the boundary conditions 
for the multigrid are given by the solution of the Poisson equation in vacuum.
Then, the total potential is constructed from the combination of the multigrid solution in
$\Omega_1$ plus the FFT solution in the rest of the space.
The full procedure is summarized in Scheme 1.

\begin{center}
\centerline{
\resizebox{3.5in}{!}{\includegraphics{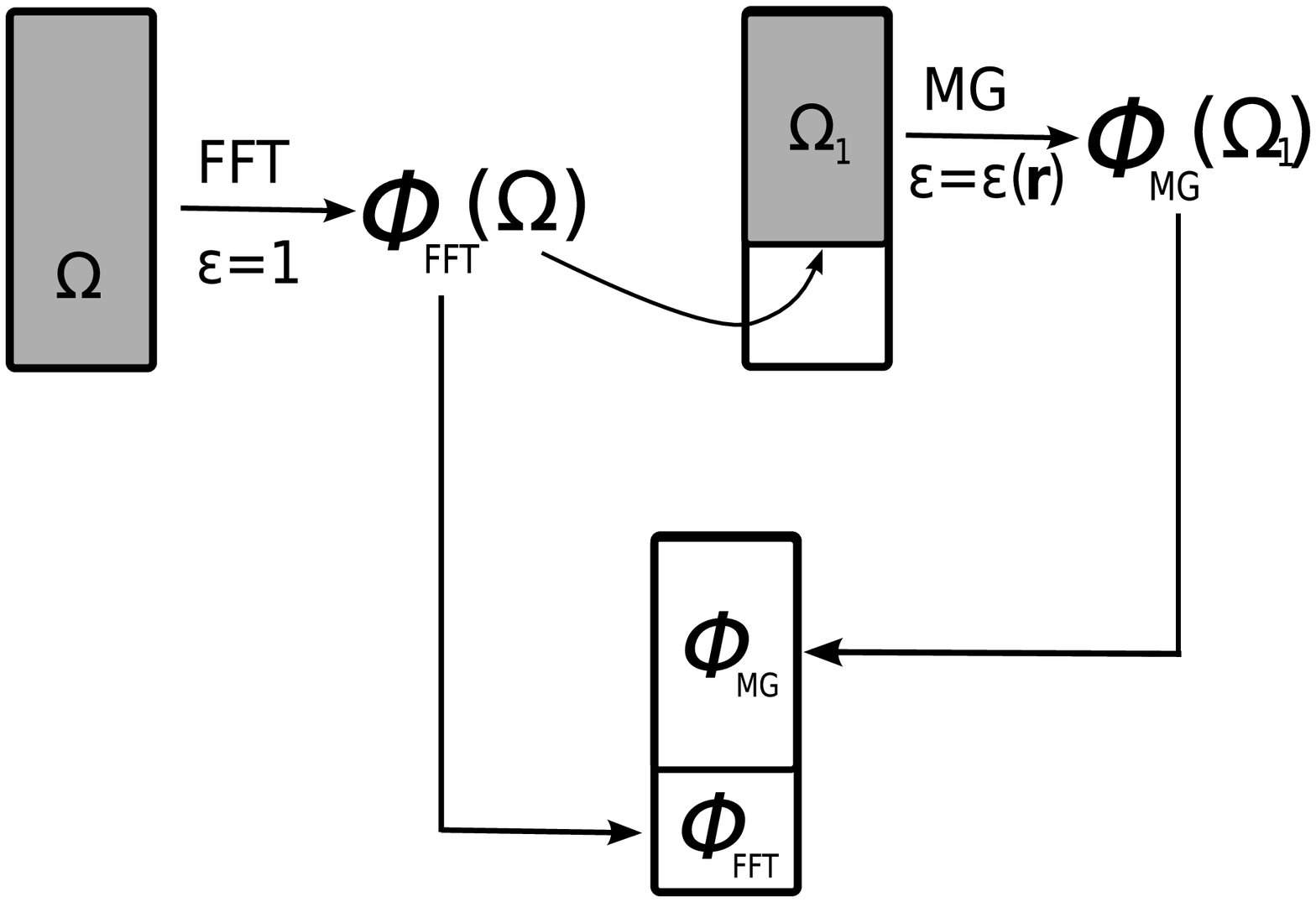}}}
Scheme 1
\end{center}

In proposing this approach, we assume the dielectric effect is mostly local and
does not perturb the electrostatic potential deep inside the slab. More formally, this implies 
that even in the presence of the dielectric medium,
there is a region in the real space grid where the self-consistent potential is exactly
the same as the potential corresponding to vacuum. 
It is therefore important that $\Omega_1$ extends
sufficiently down into the surface as to reach that region. 
With this mixed scheme, the size of the mesh to be
submitted to the multigrid routine can be reduced up to 40 $\%$.

\section{Applications: molecular dynamics at the TiO$_2$ interface}

It is well known that surface groups of most inorganic oxides ionize in solution,
exhibiting the following equilibria:
\begin{center}
M-O$^{z}$ + H$_3$O$^+$ $\Leftrightarrow$ M-OH$^{z+1}$ + H$_2$O $\Leftrightarrow$ M-OH$_2^{z+2}$ + HO$^-$
\end{center}
with $z$ the surface group charge, which can be negative, positive, or zero, depending on the
nature of the oxide.\cite{jolivet} Understanding the acid-base behavior resulting from these
equilibria is crucial in almost every application of these materials in solution.
Isoelectric points of many oxides have been known for years,\cite{crisoele} 
however, it is very difficult to probe the surface of bulk materials
or nanoparticles in solution, and most of the data collected corresponds to the average behavior of
the different surfaces exposed in a given experiment.
More recently, researchers have sought to take advantage of density functional theory to
establish the degree of dissociation and protonation at different
titania-water interfaces with an explicit representation 
of the solvent.\cite{bandura}$^-$\cite{lang2} For the reasons already discussed, such
an approach is costly and has been employed only in a limited number of cases.
In what follows, we apply our continuum solvent model to characterize the hydrated (101)
surface of the anatase structure of TiO$_2$, which is possibly the most stable and abundant.\cite{diebold}
The adsorption of H$_2$O on perfect TiO$_2$ surfaces in the gas phase has been thoroughly
investigated using both experimental and theoretical approaches.\cite{tio2_1}$^-$\cite{tio2_c}
In the case of the (101) face of anatase, there is consensus in the fact that molecular
adsorption of water is thermodynamically the most stable. Electronic
structure calculations suggest that the difference between the two possible adsorption 
modes---molecular versus dissociated---is of nearly 10 kcal/mol.\cite{tio2_a} 
We have performed calculations in
four layers slabs representing the (101) surface of the anatase structure.
As previously reported, our own calculations in vacuum summarized in Table II
show that at different water coverages, the molecular pathway
is the most favored. The same trend prevails in the presence of the solvent, although
the interaction energies with the surface turn out to be significantly lower. Table II presents
the results from geometry optimizations of the fully hydrated surface embedded in the
continuum dielectric. The observed weakening of the interaction with the oxide with respect to
vacuum is a consequence of the stabilization of the H$_2$O molecules in the polar environment,
and can be understood in terms of a competition between the bulk solvent and the
surface for the water molecules. Despite this lower affinity, the massive presence of
water from the liquid phase will displace the equilibrium toward the formation of an adsorbed
monolayer. The energy difference between the two kinds of mechanisms remains about the same as
in the gas phase, the dissociative adsorption becoming exothermic.
In solution, however, dissociation is likely to occur, controlled by the pH of the medium (see below).

\begin{figure} \centerline{
\rotatebox{-90}{\resizebox{4in}{!}{\includegraphics{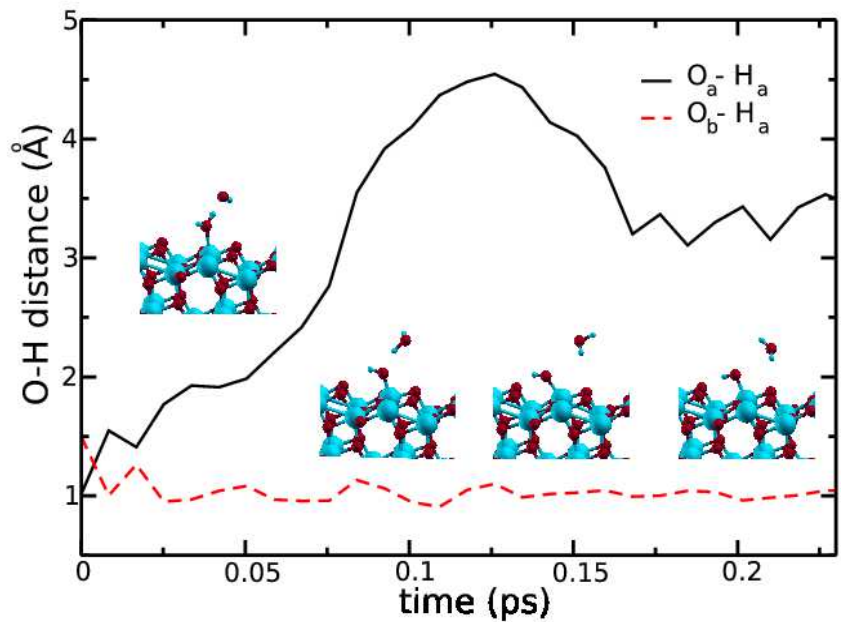}}}
}
\caption{Interatomic distances O$_a$-H$_a$ and O$_b$-H$_a$ during
a molecular dynamics simulation of a proton transfer process, from an adsorbed water molecule
to an hydroxide ion in a continuum solvent. O$_a$ and O$_b$ are
the oxygen atoms of the initially absorbed water molecule and the hydroxide, respectively, and
H$_a$ is the exchanged proton. At the beginning of the simulation the OH$^-$ group is
H-bonded to the water molecule. The geometry of the system is shown in the inset for selected time steps. 
}
\label{fig5}
\end{figure}

\begin{figure} \centerline{
\rotatebox{-90}{\resizebox{4in}{!}{\includegraphics{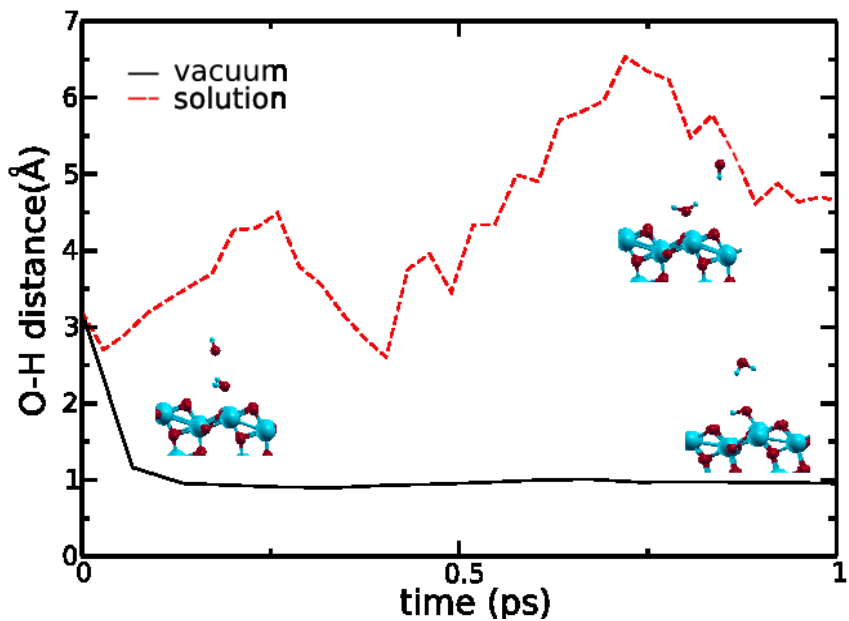}}}
}
\caption{Interatomic distance O$_b$-H$_a$ during the
molecular dynamics simulations of an adsorbed water molecule in the presence of
an hydroxide ion initially situated 4$\AA$ above the surface. O$_b$ is
the oxygen atom of the hydroxide and H$_a$ is a proton of the water molecule, which
is rapidly exchanged during the gas phase dynamics. The starting geometry is
shown on the left, while the upper and lower figures on the right depict the atomic structures after
1 ps of dynamics in vacuum and in solution, respectively.
}
\label{md2}
\end{figure}

The quantitative effect of pH on the ionization of the surface is quite difficult to assess 
from first principles simulations, since a huge supercell would be needed to have a meaningful
representation of the proton concentration in the system.
In this preliminary study, we limit ourselves to examine the proton exchange between
an adsorbed water molecule and an hydroxyl anion from the solution, using molecular dynamics at 300 K.
This computational experiment is meant to provide a qualitative picture of the abstraction of
a proton from the surface in the presence of OH$^-$, 
illustrating at the same time the performance of the continuum solvent method. 
The inset of Fig.~\ref{fig5} shows the evolution of the dynamics through a sequence of photos,
starting at an initial configuration in which an OH$^-$ group exhibits an H-bond with
an adsorbed water molecule. Early in the simulation, 
the covalent O-H bond in H$_2$O is disrupted, to leave an hydroxyl 
function on the surface and a newly formed water molecule that soon wanders around the liquid phase.
Fig.~\ref{fig5} presents the O$_a$-H$_a$ and O$_b$-H$_a$ distances, where O$_a$ and O$_b$ are
the oxygen atoms of the (initially) absorbed water molecule and the hydroxide, respectively,
and H$_a$ is the abstracted proton.
For this reaction to occur in a reasonable time as to be within reach of molecular dynamics simulations, 
the starting geometry must be chosen appropriately.
As a matter of fact, had the simulation been started from a random configuration,
the OH$^-$ anion might have explored the supercell for several picoseconds
without ever reacting with the water molecule.
In the absence of the solvent, instead, the unscreened interaction between the hydroxide and 
the surface leads to an immediate reaction. This distinctive behavior is displayed in Fig.~\ref{md2}, 
which presents the O$_b$-H$_a$ distances for two simulations, one in vacuum 
and the other in solution, started from the same geometry and with
identical computational parameters. 
The observed contrast between the two dynamics
evinces how the dielectric medium stabilizes the hydroxide in the liquid phase.
The use of this kind of methodology 
in combination with weighted importance sampling techniques (e.g., Umbrella
sampling)\cite{umbrella} could provide estimates for the acid-base equilibrium constants corresponding
to different oxide surfaces and phases. We believe this direction,
even though beyond the scope of the present work---with an emphasis 
on the methodological conception---aims to a very appealing ground for future research.

\section{Closing remarks}

We have shown that a dielectric medium defined as a function of the self-consistent charge density
provokes a strong response in the effective potential, which in solid-liquid systems
may spoil the convergence of the Car-Parrinello electronic dynamics.
Such a response can be avoided with a dielectric based on 
a non self-consistent charge, preserving in this way the potential and allowing for conservative
molecular dynamics simulations. This approach is equivalent to
have a position-dependent permittivity, and therefore a new term in the ionic forces must be
considered.

\begin{figure} \centerline{
\rotatebox{-90}{\resizebox{4in}{!}{\includegraphics{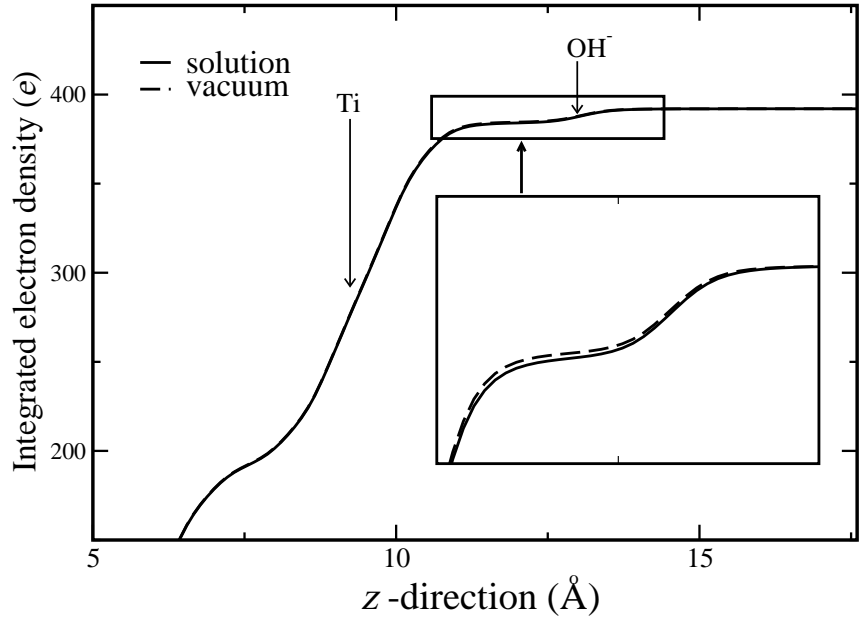}}}
}
\caption{Electron density integrated on the $xy$ plane, and displayed
as a function of the $z$ coordinate, for the hydroxyl anion situated 4$\AA$ above the
surface. The approximate positions of the upper Ti layer and of the OH$^-$ ion are
indicated with arrows. Enlargement of the interfacial region
shows a depletion of the electron density in solution with respect to vacuum.
There is an average difference of about 0.5 $e$ between the two curves, owing to the
stabilization of the electronic charge of the hydroxyl anion surrounded by the dielectric.}
\label{integratedzcharge}
\end{figure}

The methodology presented here is a powerful instrument to explore the thermodynamics and
the reactivity of surfaces and nanoparticles in solution. Replacement of the
solvent molecules by a dielectric continuum may neglect the structural features
of the liquid phase,
but it does capture the essential polarization effect of the medium. This is manifest,
for instance, in the charge of the hydroxyl group: if an additional electron is added
to a neutral system consisting of a slab plus a (distant) OH moiety in vacuum, a significant
portion of the charge flows to the solid phase. Noticeably, in the presence of the solvent,
the excess electron spontaneously localizes on the hydroxyl. This is quantitatively depicted
in Fig.~\ref{integratedzcharge}. 

A compelling application for this continuum solvent scheme, as well as a natural continuation of this 
work, would be the characterization of the adsorption energies and geometries of water 
and other species on the different surfaces of titanium dioxide.
We deem especially worthwhile the calculation of the reaction energies
for the kind of equilibria
mentioned above, e.g., Ti-OH$_2$ + OH$^- \Leftrightarrow$ Ti-OH$^-$ + H$_2$O, as a function of
the surface structure. Work in this direction is now in progress. 
At this point it should be noted that it would be not feasible to have an estimate of these quantities
without the solvation model: in the absence of the dielectric, the interaction between the
surface and the OH$^-$ ion (or between the charged slab and the water molecule)
is extremely dependent on the distance separating them, and therefore it is not possible
to establish unequivocally the energies for reactants and products. When the dielectric
is included, the long range interaction between charged and polar fragments is efficiently
screened, and the total energy of the system becomes independent of the position of the
molecule (or the ion) with respect to the slab. This property makes possible to evaluate
reaction energies on periodic surfaces in solution which could not be calculated by other means, except
perhaps with extensive molecular dynamics simulations. 
Aside from these, the scheme presented here
would be particularly useful to assess the role of the solvent in a great diversity of problems
in surface chemistry, including the effects on structure, on vibrational
frequencies, or on charge transfer phenomena, among many others.

\section{Acknowledgments}

We feel indebted to Jean-Luc Fattebert, Oswaldo Dieguez, Ismailia Dabo,
Nicola Marzari, and Patu Groisman, for precious advice and enlightening discussions.
This study has been financially supported by CONICET (PIP-5220) and by the 
Agencia Nacional de Promoci\'{o}n Cient\'{i}fica y Tecnol\'{o}gica
(PICT 06-33581). V.M.S. acknowledges CONICET for a doctoral fellowship.

\section{Appendix A: functional derivative of V$_\epsilon$}

Following Parr and Yang,\cite{parryang} the functional derivative of $E_{H}[\rho]$ with respect
to $\rho({\bf r})$, which we will denote $\frac{\delta E_{H}}{\delta \rho}$, is defined
by the relation
\begin{equation}
\lim\limits_{\lambda\to 0}\left[\frac{E_{H}(\rho+\lambda
f)-E_{H}(\rho)}{\lambda}\right] =
\int \frac{\delta E_{H}}{\delta \rho} f({\bf r}) d{\bf r}
\end{equation}
where $f({\bf r})$ is arbitrary. From the definition of $E_{H}[\rho]$ we can write
\[
2E_{H}(\rho+\lambda
f)-2E_{H}(\rho)=\int\phi_{\rho+\lambda f}(\rho+\lambda f)\,d{\bf r}-
\int \phi_{\rho}\rho\,d{\bf r} \] 
\begin{equation}
= \int\phi_{\rho+\lambda f}\,\rho\ d{\bf r} + \lambda \int
\phi_{\rho+\lambda f} \,f\,d{\bf r}\
-\int \phi_{\rho}\rho\,d{\bf r}
=\lambda \int \phi_{\rho+\lambda f} \,f\,d{\bf r} +
\int \left(\phi_{\rho+\lambda f}-\phi_{\rho}\right)\rho\,d{\bf r}.
\end{equation}
In the last term above $\rho$ can be rewritten using Eq. (3), and the resulting expression
can be integrated by parts (from now on we omit $d{\bf r}$ from the integrand for conciseness):  
\[\int \left(\phi_{\rho+\lambda f}-\phi_{\rho}\right)\rho=
-\frac1{4\pi}\int
\left(\phi_{\rho+\lambda
f}-\phi_{\rho}\right)\nabla\left(\epsilon_\rho\nabla\phi_\rho\right) \]
\[
=\frac1{4\pi}\int\nabla\phi_{\rho+\lambda
f}\,\epsilon_\rho\nabla\phi_\rho\;-\; \frac1{4\pi}\int
\nabla\phi_{\rho}\,\epsilon_\rho \nabla\phi_\rho \]
\[
=\frac1{4\pi}\int\nabla\phi_{\rho+\lambda
f} \left(\epsilon_\rho-\epsilon_{\rho+\lambda
f}\right)\nabla\phi_\rho +
\frac1{4\pi}\int\nabla\phi_{\rho+\lambda f}\,\epsilon_{\rho+\lambda
f}\nabla\phi_\rho \;-\; \frac1{4\pi}\int
\nabla\phi_{\rho}\,\epsilon_\rho\nabla\phi_\rho\]
\[
=\frac1{4\pi}\int\nabla\phi_{\rho+\lambda
f}\,\left(\epsilon_\rho-\epsilon_{\rho+\lambda
f}\right)\nabla\phi_\rho\;+\; \int (\rho+\lambda f) \phi_\rho \;-\;
\frac1{4\pi}\int \epsilon_\rho (\nabla\phi_\rho)^2.\]
To arrive to the last expression, $\epsilon_{\rho+\lambda f}$ was added and subtracted in
the third equality, and then the second term was integrated by parts.
This result can now be inserted in Eq. (21):
\[
2E_{H}(\rho+\lambda f)-2E_{H}(\rho)= \lambda \int \phi_{\rho+\lambda f} \,f
+ \frac1{4\pi}\int\nabla\phi_{\rho+\lambda
f}\,\left(\epsilon_\rho-\epsilon_{\rho+\lambda
f}\right)\nabla\phi_\rho\; + \lambda \int f \phi_\rho \]
\begin{equation}
+ \int \rho \phi_\rho - \frac1{4\pi}\int \epsilon_\rho (\nabla\phi_\rho)^2 =
\lambda \int (\phi_{\rho+\lambda f} + \phi_\rho) \,f - \frac1{4\pi}\int\nabla\phi_{\rho+\lambda
f}\,\left(\epsilon_{\rho+\lambda f}-\epsilon_\rho \right)\nabla\phi_\rho
\end{equation}
Dividing Eq. (22) by $2\lambda$, and taking the limit ($\lambda \to 0$), we get to the
final outcome:
\[
\lim\limits_{\lambda\to 0}\left[\frac{E_{H}(\rho+\lambda f)-E_{H}(\rho)}{\lambda}\right] =
\lim\limits_{\lambda\to 0} \left[ \frac{1}{2}\int (\phi_{\rho+\lambda f} + \phi_\rho) \,f 
- \frac1{8\pi}\int\nabla\phi_{\rho+\lambda f}\,\nabla\phi_\rho
\frac{\left(\epsilon_{\rho+\lambda f}-\epsilon_\rho \right)}{\lambda} \right] \]
\begin{equation}
=\int \phi_\rho \,f - \frac1{8\pi}\int\nabla\phi_\rho \nabla\phi_\rho \frac{\partial\epsilon}{\partial\rho}\,f
=\int (\phi_\rho - \frac1{8\pi}(\nabla\phi_\rho)^2 \frac{\partial\epsilon}{\partial\rho})\,f({\bf r}) d{\bf r}. 
\end{equation}
By comparison with Eq. (20), it is easy to see that 
$\frac{\delta E_{H}}{\delta \rho}=\phi_\rho - \frac1{8\pi}(\nabla\phi_\rho)^2 
\frac{\partial\epsilon}{\partial\rho}$.

\section{Appendix B: derivative of $E_H$ with respect to the ionic positions}

In the absence of a dielectric ($\epsilon= 1$), Eq. (15) can be inserted in the expression
for the Hartree energy to give
\begin{equation}
E_H=2\pi\Omega \sum_{\bf G} \frac{\left| \tilde \rho({\bf G})\right|^2}{G^2}
\end{equation}
where $\tilde\rho({\bf G})$ are the Fourier coefficients for the expansion of $\rho_{tot}({\bf r})$,
$\tilde\rho({\bf G})=\tilde\rho_e({\bf G})+
\sum_I \tilde\rho_I({\bf G}) e^{-i{\bf G R_I}}$.
Since the only explicit dependence of $\tilde\rho({\bf G})$ on $\{{\bf R_I}\}$ is through the structure
factor ($e^{-i{\bf G R_I}}$), the derivative of Eq. (24) with respect to the atomic positions is just
\begin{equation}
\frac{\partial E_H}{\partial {\bf R_I}}=
-4\pi\Omega \sum_{\bf G} i{\bf G} \left(\frac{\tilde \rho^*({\bf G})}{G^2}\right)
\tilde \rho_I({\bf G}) e^{-i{\bf G R_I}}.
\end{equation}
In the presence of a dielectric medium determined by the ionic coordinates, Eq. (24) does not hold. 
To calculate $\partial E_H/\partial {\bf R_I}$ we replace $\epsilon[\rho]$ for
$\epsilon({\bf R_I})$ in Eq. (4) and derivate:
\begin{equation}
\frac{\partial E_H}{\partial {\bf R_I}}
= \frac{1}{8\pi} \int \frac{\partial \epsilon({\bf R_I})}
{\partial {\bf R_I}}(\nabla \phi({\bf r}))^2d{\bf r}
+ \frac{1}{8\pi} \int \epsilon({\bf R_I}) \frac{\partial (\nabla \phi({\bf r}))^2}{\partial {\bf R_I}}d{\bf r}.
\end{equation}
The second term on the right member can be further developed as follows:
\[
\frac{1}{8\pi} \int \epsilon({\bf R_I}) \frac{\partial (\nabla \phi({\bf r}))^2}{\partial {\bf R_I}}d{\bf r}=
\frac{2}{8\pi} \int \epsilon({\bf R_I}) \nabla \phi({\bf r}) 
\left(\nabla \frac{\partial \phi({\bf r})}{\partial {\bf R_I}}\right) d{\bf r}= \]
\begin{equation}
-\frac{1}{4\pi} \int \nabla \cdot \left[\epsilon({\bf R_I}) \nabla \phi({\bf r})\right]
\frac{\partial \phi({\bf r})}{\partial {\bf R_I}} d{\bf r}=
\int \rho_{tot}({\bf r}) \frac{\partial \phi({\bf r})}{\partial {\bf R_I}} d{\bf r}
\end{equation}
where we have integrated by parts and used Eq. (3).
Then, it is possible to rewrite (26):
\begin{equation}
\frac{\partial E_H}{\partial {\bf R_I}}
= \frac{1}{8\pi} \int \frac{\partial \epsilon({\bf R_I})}
{\partial {\bf R_I}}(\nabla \phi({\bf r}))^2d{\bf r}
+\int \rho_{tot}({\bf r}) \frac{\partial \phi({\bf r})}{\partial {\bf R_I}} d{\bf r}.
\end{equation}
On the other hand, the derivation of 
the general expression for the Hartree energy leads to:
\begin{equation}
\frac{\partial E_H}{\partial {\bf R_I}}
= \frac{1}{2} \int \rho_{tot}({\bf r})\frac{\partial \phi({\bf r})}{\partial {\bf R_I}} d{\bf r}
+ \frac{1}{2} \int \phi({\bf r}) \frac{\partial \rho_{tot}({\bf r})}{\partial {\bf R_I}}d{\bf r}.
\end{equation}
Equating (28) and (29) we find the following relation:
\begin{equation}
\int \rho_{tot}({\bf r}) \frac{\partial \phi({\bf r})}{\partial {\bf R_I}} d{\bf r}=
-\frac{1}{4\pi} \int \frac{\partial \epsilon({\bf R_I})}
{\partial {\bf R_I}}(\nabla \phi({\bf r}))^2d{\bf r} + 
\int \phi({\bf r}) \frac{\partial \rho_{tot}({\bf r})}{\partial {\bf R_I}}d{\bf r}.
\end{equation}
Ultimately, replacing (30) into (28) we obtain the final result,
\begin{equation}
\frac{\partial E_H}{\partial {\bf R_I}}
= -\frac{1}{8\pi} \int \frac{\partial \epsilon({\bf R_I})}
{\partial {\bf R_I}}(\nabla \phi({\bf r}))^2d{\bf r}
+ \int \phi({\bf r}) \frac{\partial \rho_{tot}({\bf r})}{\partial {\bf R_I}}d{\bf r}.
\end{equation}

\newpage

\begin{table}
\caption{Solvation free energies (kcal/mol) for selected molecules and ions in water, calculated
with this model using a dielectric function $\epsilon=\epsilon(\rho)$ determined by the 
electron density,$^a$ and
$\epsilon=\epsilon(\gamma({\bf R_I}))$ determined by the atomic positions with $\beta$=2.4 (see text). 
Experimental values$^b$ and results from PCM$^c$ are also shown.}

\begin{center}
\begin{ruledtabular}
\begin{tabular}{lcccc}
 & Expt. & $\epsilon=\epsilon(\rho)$ & $\epsilon=\epsilon(\gamma)$ & PCM \\
\cline{2-5} 
H$_2$O & -6.3 & -8.4 & -8.8 & -5.4 \\
CH$_3$CONH$_2$ & -9.7 & -10.5 & -8.0 & -4.6 \\
CH$_3$NH$_3^+$ & -73 & -81.0 & -81.9 & -65.1 \\
NO$_3^-$ & -65 & -57.8 & -60.6 & -62.6 \\
Cl$^-$ & -75 & -66.9 & -68.6$^d$ & -72.6 \\
\end{tabular}
\end{ruledtabular}
\end{center}
\noindent
$^a$ From reference~\cite{jcp0}.
$^b$ From references~\cite{sol1,sol2}.
$^c$ Obtained with the Polarizable Continuum Model as implemented in
Gaussian 03.\cite{crtomasi1, cpl-cossi}
$^d$ The van der Waals radius for ionic chlorine was set equal to 2.059 $\AA$, 
from reference \cite{clradius}.
\end{table}

\begin{table}
\caption{Adsorption energies for water on the anatase (101) surface in the gas
phase and in solution, in both the molecular and dissociative configurations at
different coverages. Values are given in kcal per H$_2$O molecule.}
\begin{center}
\begin{ruledtabular}
\begin{tabular}{lcccc}
 & \multicolumn{2}{c} {Molecular} & \multicolumn{2}{c} {Dissociative} \\
 & $\theta = 0.25$ & $\theta = 1$ & $\theta = 0.25$ & $\theta = 1 $ \\
\cline{2-5}
 Gas phase & -19.4 & -17.8 & -1.3 & -7.4 \\
 Solution & - & -3.0 & - & 6.6 \\
\end{tabular}
\end{ruledtabular}
\end{center}
\end{table}
\

\newpage

\newpage

\newpage

\

\newpage

\newpage

\newpage

\begin{thebibliography}{10}
\bibitem{nota1}
Of course, the mere omission of $V_\epsilon$ would not be an acceptable solution, since the energy
conservation will be affected.
\bibitem{nota2}
The parameter $\beta$ is taken greater than 0.5. See reference~\cite{jluc1}.
\bibitem{nota3}
Multigrid methods solve elliptic partial differential equations by
applying a classic finite-differences iterative technique in meshes of different sizes
(in this case we adopt the Gauss-Seidel 
scheme). See reference~\cite{nrecipes} for an introduction, reference\cite{mgtutorial} 
or~\cite{mgmethods} for more specialized sources.
\end{thebibliography}
\end{document}